\documentclass{article}

\usepackage{natbib}

 \usepackage{graphics}
 \usepackage{graphicx}
 \usepackage{epsfig}

\usepackage{amssymb}

\begin{document}


 \title{Bayes and empirical Bayes changepoint problems}
\author{Heng Lian \\Division of Mathematical Sciences, SPMS\\Nanyang Technological University\\Singapore, 637616}

\maketitle


\begin{abstract}

We generalize the approach of \cite{liu} for multiple changepoint problems where the number of changepoints is unknown. The approach is based on dynamic programming recursion for efficient calculation of the marginal probability of the data with the hidden parameters integrated out. For the estimation of the hyperparameters, we propose to use Monte Carlo EM when training data are available. We argue that there is some advantages of using samples from the posterior which takes into account the uncertainty of the changepoints, compared to the traditional MAP estimator, which is also more expensive to compute in this context. The samples from the posterior obtained by our algorithm are independent, getting rid of the convergence issue associated with the MCMC approach. We illustrate our approach on limited simulations and some real data set.

\end{abstract}



\section{Introduction}
Change point models (CPM) have been one of the main research topics in statistics for many years. Inference on CPM is a difficult problem because it is an irregular model in the sense that the number of parameters in the model is not a priori fixed, thus traditional likelihood theory does not directly apply. In this paper, we aim to develop a Bayesian approach to CPM and investigate its computational properties.

Traditional approaches to CPM are frequentist in nature. Difficulties associated with the frequentist approach include inferences on the change points as well as inferences on the parameters within each single segment. The difficulties result from the fact that the number of parameters in the model increases with the number of changepoints. The frequentist approach to inferences on the number of changepoints consists mainly of adapting hypothesis testing framework for testing $i$ changepoints against $i+1$ changepoints. This approach is not only difficult to analyze but is inelegant mathematically as well.  A large portion of the literature also only focuses on binary segmentation, that is, considers only the case of at most one changepoint. The Bayesian approach is used in \cite{raftery}, but it is limited to the case of only one changepoint, and iterative binary segmentation approach is used in \cite{yang}, where Bayes factor is used to decide whether to continue segmenting the subsequence.  One approach conceptually similar to ours is \cite{green}, where the author used reversible-jump MCMC for inferences on the segmentation. That paper also demonstrated clearly that under the Bayesian framework, inferences are automatic once the model is specified. Maybe surprisingly, the posterior for Bayesian CPM can be computed in closed form with polynomial time complexity as demonstrated in \cite{liu} for DNA sequence segmentation. This approach can be generalized to continuously distributed observations (as shown in this paper) by specifying conjugate priors. This model is much more flexible than previous approaches in that we can easily switch to other observational distributions, consider correlations among observations, and use higher-order polynomial model within each segment. We acknowledge that other approaches might also be extensible to these situations with further work, but we think the hierarchical Bayesian approach is simpler for adapting to different situations once the basic framework is laid out and conceptually more elegant in dealing with different applications. The estimation and inferential procedures can be made to be fully automatic with little or no human intervention.

In this paper, we focus on the piece-wise constant model with observations following a normal distribution. It is straightforward to extend the model to other observational distribution, which we do not pursue here for both simplicity and specificity. Our Bayesian model can be described as follows. First, we have a prior for the segmentation together with the means and variances for each segment. These hidden parameters (means and variances for each segment) are drawn independently from a conjugate prior. Finally, the observations are generated independently from the normal distribution for each segment.

Section 2 sets up the Bayesian model, details the recursion used for posterior computation and specification of the hyperprior, deals with missing data and some numerical issues. Section 3 demonstrates the effectiveness of our model with simulation and application. We end with a discussion in Section 4.

\section{Hierarchical Bayesian model}
\subsection{Specification of the model}\label{sec:model}
We make use of a hierarchical model that uses a continuous mixture of normals to model the values within each segment. That is, we put a prior on the mean and variance of the observations for the segments. It turns out that the hidden parameters consisting of means and variances of our observations can be integrated out analytically, and so can the location of the changepoints. This observation implies our model has a computational complexity which is quadratic in the length of the sequence.

Our model is based on the model in \cite{liu}. The modifications and extensions we
made to that model include the following:

1. We extend the model to the case where observations are modeled as
continuous random variables. This change essentially imposes no extra technical
difficulty, if conjugate priors are used as in the discrete model, which makes
explicit integration possible.

2. In \cite{liu}, the hyperparameters are assumed to be given, which may be unsatisfactory from a statistical point of view. We propose a principled method to estimate these hyperparameters within the empirical Bayes (EB) paradigm
combined with a stochastic version of the EM algorithm, when there are multiple sequences so that estimation is feasible. We show that an essential step required by the EM algorithm is made possible by utilizing the forward recursion and backward sampling approach. We also propose a simple procedure for the automatic choice of the hyperparameters when only one sequence is available.

3. Data with incomplete observations are frequently encountered in practice. The hidden variables, which include means and variances for each segment and location of the changepoints, can also be considered as missing data. In fact, a large amount of work has been done to address the difficulty in inference arising from the existence of hidden variables, which are regarded as missing data, including the EM algorithm which is specifically designed for this purpose. In this work, we use missing data to refer to the case that some regions of the sequence are totally unobserved, producing gaps in our data. Many well-known algorithms like imputation can be used to solve this problem. We show that within our framework, the inferences can proceed without recourse to the imputation procedure. This is done by using the most common trick of Bayesian inference -- integrating as much as you can to get rid of all variables that one considers as nuisances. In contrast, the most frequently adopted strategy is to just ignore the missing observations.

Specifically, the model can be described as follows. For a sequence $y=\{y_1,y_2,\ldots, y_n\}$, suppose the maximum number of changepoints is $k_{max}$, which is specified beforehand. A particular segmentation can be denoted by $A =\{c_1,c_2,\ldots,c_k\}, 1\le c_1<c_2<\cdots <c_k=n $, where $n$ is the
length of the sequence, and $c_i$ is the $i$th changepoint. For purpose of concreteness, with two
neighboring segments, the last observation of the first segment will be considered as
the changepoint between the two. Note also that we set $c_k=n$, so the number of
changepoints is the same as the number of segments with our notation.

The recursion marginalizing over changepoints which we will present in the following is an extension of the recursion used by \cite{liu}. Marginalization over segmental means and variances is achieved by integrating over segmental means and variances in the hierarchical models. These marginalization steps greatly reduce dimensionality of the space of the unknowns.

We put a uniform prior on the number of changpoints and assume all segmentations with exactly $k$ changepoints are equally likely. That is,
\begin{eqnarray}\label{aprior}
&&p(k)=1/k_{max},\nonumber \\
& \mbox{ and } & p(A|k)=1/{n-1\choose k-1} ,\mbox{ if } A \mbox{ has $k$ changepoints.}
\end{eqnarray}

Given the segmentation $A$, the mean and variance for each segment is generated
from normal-inverse-$\chi^2$  distribution
\begin{eqnarray}\label{conj}
&&\mu_i|\sigma_i^2,A\sim N(\mu_0,\frac{\sigma^2}{k_0})\nonumber\\
&&\sigma_i^2|A\sim Inv-\chi(\nu_0,\sigma_0^2), \; i=1,2,\ldots,k
\end{eqnarray}
Those parameters with subscript $0$ are hyperparameters that need to be specified
or estimated from the data. We will refer to those $\mu_i$ and $\sigma_i^2$ above as hidden parameters,
and the segmentation $A$ itself as the hidden variable.

In our model, each segment has a different mean and variance associated with it, and these means and variances
are independently drawn from the hyperprior. With segmentation $A$ and $\mu_i, \sigma_i$ given for each segment, the observations are naturally modeled as normal with given mean and variance:
\[
y_{c_i+1:c_{i+1}}|\mu_i,\sigma_i^2,A\stackrel{iid}{\sim} N(\mu_i,\sigma_i^2),
\]
where we used the notation $y_{i:j}=\{y_i,y_{i+1},\ldots,y_{j-1},y_j\}$.  Note (\ref{conj})
is exactly the same conjugate prior that is used in \cite{gelman} which makes analytic integration
possible.

In summary, our Bayesian model can be described as follows. First, we have one prior on the hidden parameters -- the mean and variance for each segment. We also have a simple prior on the possible segmentations. Hidden parameters
for each segment are drawn independently from the hyperprior conditioned on the segmentation. Finally, the observations are generated independently from the normal distribution for each segment.

\subsection{Recursion and inferences}\label{sec:recursion}
The hyperparameters $\Theta=\{\mu_0,k_0,\nu_0,\sigma_0^2\}$ are assumed for now to be known. Then the main goal of inference is to get at the posterior distribution $p(A|y)$. In the following, we omit explicitly writing down the dependence of the probability distribution on the hyperparameters $\Theta$. We show the difficulty of computing the posterior distribution next and propose a dynamic programming recursion for solving it.

By the Bayes formula, we have $p(A|y) = p(y|A)p(A)/\sum_{A'}p(y|A')p(A')$, where $p(y|A)$
is the likelihood in which the hidden parameters are integrated out. By the independence
of $y_i$ in different segments conditioned on the hidden parameters, we have
\[p(y|A)=\prod_{i=0}^{|A|-1}p(y_{c_i+1:c_{i+1}}),\]
where $|A|$ is the number of segments in the segmentation $A$ and $c_0=0$ by convention. Each term in the product
can be analytically evaluated thanks to the conjugate prior used:
\begin{eqnarray}\label{integrate}
p(y_{a:b})&=&\int\int \frac{1}{(\sqrt{2\pi\sigma^2})^{b-a+1}}e^{-\frac{\sum_i(y_i-\mu)^2}{2\sigma^2}}\frac{1}{\sqrt{2\pi\frac{\sigma^2}{k_0}}}e^{-\frac{k_0(\mu-\mu_0)^2}{2\sigma^2}}\cdot\nonumber\\
&& \frac{(\frac{\nu_0}{2})^{\nu_0/2}}{\Gamma(\frac{\nu_0}{2})}\sigma_0^{\nu_0}\sigma^{-2(\frac{\nu_0}{2}+1)}e^{-\frac{\nu_0\sigma_0^2}{2\sigma^2}}d\mu d\sigma^2
\end{eqnarray}

The part that gives us trouble in computing $p(A|y)$ is the summation over all different segmentations. Evaluation of this sum in a brute-force manner is computationally prohibitive since the number of possible segmentations increases with $n$ like $n^{k_{max}}$. This is computationally intensive with even a sequence of reasonable length $n$ with typical choice of $k_{max}$. This sum is sometimes called partition function which is the normalizing constant in general graphical models. Approximate computation of the sum can be achieved by sampling schemes that do not require knowledge of the normalizing constant. Importance sampling (IS) or Markov Chain Monte Carlo (MCMC) can be used, but it is difficult to come up with proposal distributions that is efficient enough for our purpose due to the high dimensionality of $A$. Fortunately, a dynamic programming recursion similar to the one used in \cite{liu} can reduce the complexity to $O(n^2k_{max})$ as we explain in the following.

We denote by $p(j:i,k)$ the prior probability that $y_{j:i}$ consists of $k$ segments with $j$ the first observation of the $1$st segment, and $i$ the last observation of the $k$th segment. For simplicity, $p(1:i,k)$ is also written as $p(i,k)$. Let $p(j,k-1|i,k)$ be the conditional probability that the previous changepoint is at position $j$ given the $k$th changepoint is at position $i$. Similarly, $p(y_{i:j}|i:j,k)$ denotes the probability conditioned on the event that the subsequence from $i$ to $j$ has $k$ segments, also abbreviated as $p(y_{i:j}|k)$. Then we have

\begin{eqnarray}\label{rec}
& &pr(y_{1:i}| 1:i \mbox{ has $k$ changepoints} \mbox{ ($k$ segments) } )\nonumber\\
&=&\sum_{j<i}pr(y_{1:i}, \mbox{ last changepoint before $i$ is at } j |i, k)\nonumber\\
&=&\sum_{j<i}p(j,k-1|i,k)p(y_{1:j}|k-1)p(y_{j+1:i}|1)\nonumber\\
&=&\sum_{j<i}\frac{{j-1\choose k-2} }{{i-1\choose k-1}}p(y_{1:j}|k-1)p(y_{j+1:i}|1)\nonumber\\
\end{eqnarray}
where on the third line above we used the independence of observations for different segments given the segmentation. All of the probabilities above should be understood as being conditioned on the hyperparameters, which we omit for simplicity in notation.

The base case when $k=1$ for the above recursion can be obtained by integrating out the hidden variables:
\begin{eqnarray*}
p(y_{i:j}|1)=\int p(y_{i:j}|\mu,\sigma^2)p(\mu,\sigma^2|\mu_0,k_0,\nu_0,\sigma^2_0)\,d\mu d\sigma^2
\end{eqnarray*}
The integration can be done analytically since we used the usual conjugate prior in the model.

After $p(y|k)$ is computed by the recursion, we can compute the inverted probability $p(k|y)$ using Bayes rule:
\[ p(k|y_{1:n})\propto p(k)p(y_{1:n}|k)\]
Almost all of the desired probabilities of interest can be easily obtained. For example, the marginal likelihood is just
 \[ p(y_{1:n})=\sum_{k=1}^{k_{max}} p(k)p(y_{1:n}|k)\]

We can also compute the marginal probability that a changepoint will occur at $j, (1\le j<n) $:
\begin{eqnarray}
&&p(c_k=j \mbox{ for some } k |y)\nonumber\\
&=&\frac{1}{p(y_{1:n})}\sum_{1\le k<\kappa\le k_{max}}p(j,k)p(n-j,\kappa-k)p(y_{1:j}|k)p(y_{j+1:n}|\kappa-k)\label{eqn:marginal}
\end{eqnarray}
Using the marginal likelihood, we can also compute the  probability of a specific segmentation by $p(A|y)=p(A)p(y|A)/p(y).$ Note here we only computed the probability of a given segmentation, which does not give us a sense of which $A$ has high probability. The problem is the same as before --- we simply have too many possible segmentations and it is not feasible to compute the probabilities for all of them.

However, we can obtain exact and independent samples from the posterior distribution of the segmentations given the data. First we draw the number of segments $k$ from the posterior $p(k|y)$, which we have derived above, and set $c_k=n$. Then we can recursively sample backwards from the following distribution:
\begin{eqnarray}
&&p(c_{k-1}=j|y_{1:n},c_k=i)\nonumber\\
&\propto& p(y_{1:j}|k-1)\cdot p(y_{j+1:i}|1)p(j,k-1|i,k) \label{eqn:backward}
\end{eqnarray}
The above derivation used Bayes rule, that the conditional probability is proportional to the joint probability.

This forward recursion  backward sampling procedure is reminiscent of the forward-backward algorithm for hidden Markov model (HMM), in which the forward step computes the probability of past observations summing over all previous states, and the backward step computes the probability of future observations given the current state. This is also similar to the Viterbi algorithm, where the forward step is used to find the optimal value of the objective function, keeping track of the previous optimal state, and then backtracing is used to find the optimal states that lead to the optimal value.

If a single estimator is desired, the most frequently used one is  the maximum a posteriori  estimate (MAP). Using dynamic programming, MAP estimate can be obtained by recursion. We let $V(i,k)=\max p(c_1,..., c_{k-1}|c_k=i,y_{1:n})$, the maximum probability of the configurations that can be obtained for previous changepoints conditioned on the $k$th changepoint. Given the location $j$ of the $(k-1)$th changepoint, $V(i,k)$ is independent of the  information  contained in the segments $1,\ldots,k-2$, since the maximum probability configuration up to position $j$ is summarized by $V(j,k-1)$. So we have the following recursion:
\begin{eqnarray}
V(i,1)&=&1\nonumber\\
V(i,k)&=&\max_{j<i}p(c_{k-1}=j|c_k=i,y)\cdot V(j,k-1), \,k>1\label{eqn:map}\\
\phi(i,k)&=&\arg \max_{j}p(c_{k-1}=j|c_k=i,y)\cdot V(j,k-1), \,k>1\nonumber
\end{eqnarray}
and the optimal segmentation can be found by backtracing. First the optimal number of changepoints is determined by
\[
 k=\arg\max_{k}V(n,k)\cdot p(k|y_{1:n})
\]
Setting $c_k=n$, we recursively find the previous changepoint:
\[
 c_{k-1}=\phi(c_k,k)
\]
Note that the probability $p(c_{k-1}=j|c_k=i,y)$ used in the recursion (\ref{eqn:map}) is exactly what we used when sampling back.

Even if $k_{max}$ is considered to be a fixed constant, we need to compute $p(c_{k-1}=j|c_k=i,y)$ for $i=1,\ldots,n, j<i$. So maximum a posteriori estimate takes another $O(n^2)$ computation time after the forward recursion. A simpler method to get at the posterior distribution is to make use of the samples drawn. Drawing one sample only takes time $O(n)$ as can be seen from the derivation (\ref{eqn:backward}) above. \cite{ding} argues forcefully about the advantage of using samples to characterize the posterior distribution for their RNA structure prediction problem.

The samples can also be used to approximate the marginal probability of changepoint locations in equation (\ref{eqn:marginal}). We can simply count the fraction of samples which has a changepoint at certain positions. In equation (\ref{eqn:marginal}), $p(y_{j+1:n}|\kappa-k)$ is not directly available after forward recursion, which only computed $p(y_{1:i}|k)$ for the subsequences starting  from index $1$, so another ``backward recursion" will be required to obtain this probability. Thus the sampling approach is much easier to implement.
One advantage of using samples is that we immediately get a sense of the uncertainty in segmentation by examining different samples from the posterior.

The hidden variables, mean and variance for each segment, can also be estimated. Given a sampled segmentation, the hidden parameters can be estimated for each segment independently. Given one segment with length $l$, the prior on the hidden parameters  is the normal-inverse-$\chi^2$ distribution. The posterior distribution of the mean and variance is well known, which is in the same normal-inverse-$\chi^2$ family with updated parameters (\cite{gelman}):
\begin{eqnarray}
\mu&=&\frac{k_0}{k_0+l}+\frac{l}{k_0+l}\bar{y}\nonumber\\
k&=&k_0+l\nonumber\\
\nu&=&\nu_0+l\label{eqn:update}\\
\nu{\sigma^2}&=&\nu_0{\sigma^2_0}+\sum_i(y_i-\bar{y})^2+\frac{k_0l}{k_0+l}(\bar{y}-\mu_0)^2\nonumber
\end{eqnarray}

\begin{figure}[t]
\begin{minipage}[b]{1.0\linewidth}
  \centering
  \centerline{\includegraphics[width=11.5cm]{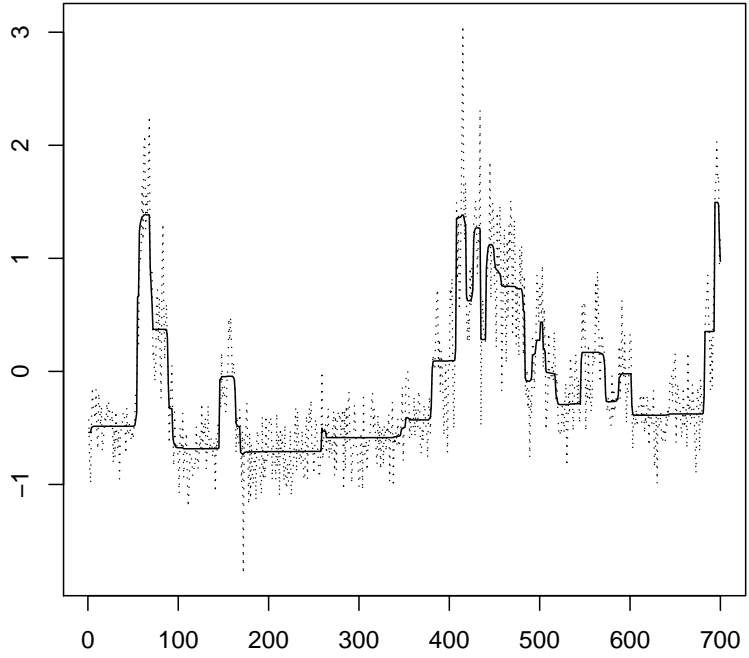}}
\end{minipage}
\caption{A data sequence (dotted) with the posterior mean of the hidden parameter $\mu$ (solid) computed from 500 samples}
\label{fig:postmean}
\end{figure}

Fix a position $i$ in the sequence. After $N$ samples are drawn, the posterior for the hidden parameters at position $i$ is a mixture of $N $  normal-inverse-$\chi^2$ distributions, with each component corresponding to one sample:
\[
\frac{1}{N}\sum_{n=1}^N N-Inv-\chi^2(\mu,\sigma^2|\mu^{(n)},k^{(n)},\nu^{(n)},{\sigma^2}^{(n)})
\]
where $\mu^{(n)},k^{(n)},\nu^{(n)},{\sigma^2}^{(n)}$ are the values computed as in (\ref{eqn:update}) using the segment that contains $i$ in the $n$-th sample. We can use the mean of this mixture distribution as a summary statistics computed from multiple samples. Figure \ref{fig:postmean} shows the mean of $\mu_i$, which looks like a smoothed version of the original data. For this visually complex sequence, changepoint model is best seen as a tool for function approximation or denoising. One possible usage of these estimated hidden parameters is to check model fit by computing the standardized residuals at each position.

This simple model can be extended somewhat by taking into account the length constraint. The constraint on the segment length can be represented as an interval $[l, u]$, where $l\ge 1$ is the lower bound and $u\le n$ is the upper bound for the length of a segment. These bounds might come from expert opinion. Putting a lower bound might also make the segmentation result more robust to outliers.

With length constraints on the segments, the simple combinatorial counting can no longer be used. A recursion is required to count the number of possible segmentations. Let $S(i,k)$ be the number of possible segmentations of a sequence of length $ i $, with $k$ changepoints ($k$ segments). $S(i,k)$ can be computed using the following recursion:
\begin{eqnarray*}
S(i,1)&=&\left\{\begin{array}{ll}
    1 &\mbox{ if } i\in [l,u]\\
    0 & \mbox{otherwise}
    \end{array}\right.\\
S(i,k)&=&\sum_{j<i} S(j,k-1)S(i-j,1)\\
\end{eqnarray*}
The main recursion now becomes:
\begin{eqnarray*}
& &pr(y_{1:i}| 1:i \mbox{ has } k \mbox{ changepoints } ( k \mbox{ segments)} )\\
&=&\sum_{j<i}p(j,k-1|i,k)p(y_{1:j}|k-1)p(y_{j+1:i}|1)\\
&=&\sum_{j<i}\frac{S(j,k-1)S(i-j,1)}{S(i,k)}p(y_{1:j}|k-1)p(y_{j+1:i}|1)\\
\end{eqnarray*}
Note that we can define $\hat{p}(y_{1:i}|k):=S(i,k)p(y_{1:i}|k) $ ($\hat{p}$ is not probability), so the recursion can be done in terms of $\hat{p}$, which becomes simply
\[
\hat{p}(y_{1:i}|k)=\sum_{j}\hat{p}(y_{1:j}|k-1)\hat{p}(y_{j+1:i}|1)
\]

Actually, the recursion (\ref{rec}) can also be transformed into a simpler form by defining $\hat{p}$ appropriately. With this definition of $\hat{p}$, other formulae also become simpler. For example, the recursive backward sampling becomes
\begin{eqnarray*}
p(c_{k-1}=j|y_{1:n},c_k=i)&\propto&\hat{p}(y_{1:j}|k-1)\cdot\hat{p}(y_{j+1:i}|1)
\end{eqnarray*}

\subsection{Estimation of the hyperparameters}\label{sec:eb}
We have up to now assumed that the hyperparameters $\Theta$ are fixed in advance. In some simple cases, those hyperparameters can be specified beforehand by an expert who has some  idea about the magnitude of these parameters due to previous experience. Although this sounds like a simple task, prior elicitation is the most important and time consuming part of Bayesian analysis. Good priors are usually difficult to obtain, and the poorly chosen prior  will usually bias the inference, although this practical issue does not seem to concern theoretical statisticians.  Empirical Bayes (EB) is a principled method to estimate hyperparameters in a Bayesian model, and we will discuss it in this subsection.


For our hierarchical Bayesian model, the marginal probability of the sequence is $p(y|\Theta)$, after summing over segmentations and integrating out hidden parameters. This marginal probability is computed by forward recursion in section \ref{sec:recursion}, which takes $O(n^2)$ time. Although dynamic programming recursion has provided an efficient way to compute this for fixed hyperparameters, it does not solve our problem. If we use some numerical software package to optimize this objective function, a large number of evaluations are probably required to find the optimum, which is too costly in terms of computation time. Also, many efficient numerical optimization algorithms require the gradient as an input, which is also difficult to calculate for our problem. In the following, we will see an approach that makes the parameter estimation problem possible.

As discussed above, we cannot directly perform maximization of the marginal likelihood to get an estimate for the hyperparameters since for each fix set of hyperparameters, evaluating the likelihood would require performing the recursion from scratch, which is too expensive since the optimization procedure will require many such evaluations. A natural approach to avoid this is to use the expectation-maximization (EM) algorithm and regard the segmentation as the missing variables. The EM algorithm is an efficient method dealing with missing data problems. In the E-step, it computes the sufficient statistics of $A$ in the complete model $\log p(A,y|\Theta)$ under the distribution $p(A|y,\Theta^{(old)})$, where $\Theta^{(old)}$ is the hyperparameters obtained from the previous iteration. In the M-step, we maximize over $\Theta$ the expectation  of the log-transformed complete likelihood conditioned on the old hyperparameters  to obtain the new hyperparameters:
\[ \Theta^{(new)}=\arg\max_{\Theta} E[\log p(A,y|\Theta)|y,\Theta^{(old)})]\]
So the EM algorithm works by maximizing $E[\log p(A,y|\Theta)|y,\Theta^{(old)})]$ with respect to $\Theta$, given the previous estimate $\Theta^{(old)}$, and then replace it in the next iteration with the new $\Theta$ obtained by maximization in the previous step. The EM algorithm was shown to increase the likelihood in each iteration and will converge to the (local) maximum (\cite{wu}).

The expectation $E[\log p(A,y|\Theta)|y,\Theta^{(old)})]$ can be written out as
\[\sum_A p(A|y,\Theta^{(old)})\log p(A,y|\Theta)\]
Again, we are confronted with the problem of having to deal with the task of summing over all possible segmentations. To avoid this, we use a variant of EM called Monte Carlo EM (MCEM). The difference between MCEM and ordinary EM is that we use samples from the distribution $ p(A|y,\Theta^{(old)}) $ to approximate the sum. That is, we replace the summation above with samples drawn from the posterior using hyperparameters obtained from the last iteration:  \[
E[\log p(A,y|\Theta)|y,\Theta^{(old)})]\approx \frac{1}{N}\sum_{n=1}^N\log p(A^{(n)},y|\Theta)  \]
where $\{A^{(n)}\}_1^N $ are samples from $p(A|y,\Theta^{(old)})$. But now, we can no longer guarantee that the algorithm will return a (local) maximum of the likelihood, or even that it will converge at all, although it generally works quite well in practice, and arbitrary accuracy can be obtained with a large number of samples.

The sum $\sum_n\log p(A^{(n)},y|\Theta) $ can be written as
\begin{eqnarray*}
\sum\log p(A^{(n)},y|\Theta) &=&\sum_n \log [p(A^{(n)})p(y|A^{(n)},\Theta)]\\
&=&\sum_n\sum_{(i,j)\in A^{(n)}}\log p(y_{i:j}|1,\Theta)+\mbox{constant}
\end{eqnarray*}
where the second sum is over all segments of $A^{(n)}$. Notice in the computation of the above, we only need to evaluate the integration in (\ref{integrate}) which has a closed-form formula, no recursions are required. Recursions are required for each iteration of the EM algorithm. Empirically, we observe 10 iterations seem to be sufficient for EM to reach convergence in our examples.

The estimation of hyperparameters is feasible only when we have multiple sequences or a long sequence with enough number of segments. If this is not true, the following data dependent choice for the hyperparameters can be used and works well in our experience:
\begin{eqnarray}\label{autoprior}
&&\mu_0: \mbox{ mean of the data}\nonumber\\
&&k_0: \,0.01\nonumber\\
&&\nu_0: \,3\\
&&\sigma_0^2: \mbox{ variance of the data}\nonumber
\end{eqnarray}
These choices are the same as in \cite{rafterymixture} with the same reasoning, except that we find empirically using variance of the data for $\sigma_0^2$ works better in our problem.

\subsection{Missing observations}\label{sec:missing}
Our model also yields a principled means to span the missing observations. Intuitively, when the gaps are small, we expect observations at the beginning and end of the gaps belong to the same segment, but this correlation across missing observations is unlikely to span long gaps.

Let's suppose that we make no observations on a subset $I\subseteq\{i,i+1,\ldots, j\}$ of a segment $[i,j]$. In the forward recursion, we need to compute the probability $p(y_{obs}|1)$, where $y_{obs}$ is the observed values within $[i,j]$, which can be written as the integral over the missing observations
\[
p(y_{obs}|1)=\int p(y_{i:j}|1)dy_{mis}
\]
We can simply ignore the missing observations in the computation of $p(y_{i:j}|1)$, the missing observation is taken into account only when we count the number of segmentations using the prior (\ref{aprior}).
This sounds more complicated than it really is. For example, the recursion and sampling can be done exactly as before by setting $p(y_{j_1:j_2}|1)=1$ if all values inside $[j_1,j_2]$ are missing.

We want to point out that in fact we made an important assumption which may not be obvious to the reader. Without any assumption, we have for all observed data $y_{obs}$ inside the $i$th segment
\[
p(y_{obs},\mathbf{r})=\int p(y_{obs},y_{mis}|\mu_i,\sigma_i^2)p(\mathbf{r}|y_{mis},y_{obs}) p(\mu_i,\sigma_i^2|\Theta)d\mu_i\sigma_i^2 dy_{mis}
\]
where $\mathbf{r}$ is a vector containing binary variables indicating the positions of the missing observations. Because of the integration over $y_{mis}$, this will potentially introduce complicated correlations between components of $y_{obs}$. So the integration over $\mu_i$ and $\sigma_i^2$ generally cannot be computed. If we in addition assume that $p(\mathbf{r}|y_{mis},y_{obs})=p(\mathbf{r}|y_{obs})$, that is all the information contained in $\mathbf{r}$ can be derived from the observed data, then it can be easily seen that integration over $y_{mis}$ will not introduce any correlation between the observed data given $\mu_i$ and $\sigma_i^2$, so $p(y_{obs})$ can be computed by ignoring the missing data. This assumption is called Missing At Random (MAR) (\cite{little}). This is the working assumption under which our approach works.

As a simple illustration, we simulated the observations as follows:
\begin{eqnarray*}
&&y_i\sim N(-1,1), 1\le i\le 100\\
&&y_i\sim N(-0.6,1), 101\le i\le 150\\
&&y_i\sim N(1,1),151\le i\le 250
\end{eqnarray*}
Note the changepoint between the first two segments are more difficult to detect. Then we artificially insert some missing observations between the first two segments, so the observed data are:
\begin{eqnarray*}
&&y_i\sim N(-1,1), 1\le i\le 100\\
&&y_i\sim N(-0.6,1), 201\le i\le 250\\
&&y_i\sim N(1,1),251\le i\le 350
\end{eqnarray*}
We generate 100 sequences of data from the first situation, and change the index of the observations to get 100 sequences with missing observations. We use the simple choice for the hyperparameters stated at the end of section \ref{sec:eb}. For the first situation without missing observations, the algorithm detected the changepoint for 14 of the sequences (we decide that the changepoint is found if there is a changepoint within $3$ observations of the true one). While for the second situation, the algorithm detected the changepoint for $55$ of the sequences. This simple simulation illustrated that sometimes whether we take into account the missing observation makes a big difference. The mechanism for dealing with missing data will not be used further in the rest of the paper.

\subsection{Numerics}\label{sec:num}
Direct implementation of our algorithm only works for sequences of a couple of hundred observations long. For  longer sequences, the underflowing or overflowing of float numbers should be properly dealt with. In computing the MAP estimate. This problem is easily solved by using log probability and thus transforms the products into sums. This does not solve our problem in forward recursion and some other steps of the algorithm because sums are mixed together with products in the recursion.


In the following we explain our approach to solve this problem. Although we suspect that many other researchers have used this technique, we cannot find it documented in the literature. A different strategy used for HMM, well documented in \cite{rabiner}, does not seem to be adaptable to our situation.

We illustrate our approach with the computation of (\ref{rec}). The main idea is actually the same as in computing the MAP. In the computer memory, the probability $p(y_{1:i}|k)$ is stored as its log $lp(y_{1:i}|k):=\log p(y_{1:i}|k)$. Computation of $\log p(y_{1:i}|k)$ by recursion (\ref{rec}) should  proceed with care. We have the recursion:
\[
p(y_{1:i}|k)=\sum_{j<i}a_{ijk}p(y_{1:j}|k-1)p(y_{j+1:i}|1),
\]
where $a_{ijk}={j-1\choose k-2}/{i-1\choose k-1}$.

Suppose the log of $p(y_{1:j}|k-1)$ and $p(y_{j+1:i}|1) $ in the above have been stored in the memory, we might be tempted to write everything in the exponential from, in which case we have
\[
p(y_{1:i},k)  =\sum_{j}exp\{lp(y_{1:j}|k-1)+lp(y_{j+1:i}|1)+\log a_{ijk}\}
\]
But taking exponential directly does not work --- the reason that we choose to store $lp(y_{1:i}|k)$ instead of $p(y_{1:i}|k)=exp\{lp(y_{1:i}|k)\}$ in the first place is that computation of $exp\{lp(y_{1:i}|k)\}$ might lead to overflow or underflow. This is still true with $exp\{lp(y_{1:j}|k-1)+lp(y_{j+1:i}|1)+\log a_{ijk}\}$. Here is the trick to avoid this. Let $c=\max_{j}\{ lp(y_{1:j}|k-1)+lp(y_{j+1:i}|1)+\log a_{ijk} \}$, so
\begin{eqnarray*}
p(i,k)&=&\sum_{j}exp\{lp(y_{1:j}|k-1)+lp(y_{j+1:i}|1)+\log a_{ijk} \}\\
&=&(\sum_{j}exp\{lp(y_{1:j}|k-1)+lp(y_{j+1:i}|1)+\log a_{ijk}-c\})exp\{c\}
\end{eqnarray*}
Now the numbers in the exponent $lp(y_{1:j}|k-1)+lp(y_{j+1:i}|1)+\log a_{ijk}-c$ is nonpositive, with the largest among them exactly 0. Although after taking exponential, there might be some positive terms that will become 0 due to underflow, we can be sure that these terms do not have significant contributions to the sum, because the sum is dominated by terms that are close to 1. In this way we separate the terms that are the dominating ones from the others, and these terms are computed with high precision. The probability $p(y_{1:i}|k)$ will finally be stored in memory as
\[
\log(\sum_{j}exp\{lp(y_{1:j}|k-1)+lp(y_{j+1:i}|1)+\log a_{ijk}-c\})+c
\]

\subsection{Markov dependence}
Our basic model in Section \ref{sec:model} assumes that each segment has its own mean and variance, and observations within one segment are i.i.d. given these hidden parameters. In this subsection, we will present an extension that takes into account higher order Markov dependency within one segment. Here we only consider one particular Markov model that is called autoregressive model which has been studied a lot in parametric time series theory. A good introductory book on the classical theory for mathematically inclined readers is \cite{brockwell}, while the more recent book  \cite{fan} focuses more on the nonparametric aspect of the theory.  An autoregressive model of order $p$, $AR(p)$, is defined as
\begin{eqnarray}
y_t=\beta_1 y_{t-1}+\cdots+\beta_p y_{t-p}+\epsilon_t\label{eqn:ar}
\end{eqnarray}
where in general $\epsilon_t$ is a white noise process $WN(0,\sigma^2)$, i.e.
\[
E(\epsilon_t)=0, \, Var(\epsilon_t)=\sigma^2, \, \mbox{ and }\, Cov(\epsilon_t,\epsilon_s)=0, \mbox{ for all }  t\neq s
\]
We assume the more strict condition that $\epsilon_t$ are i.i.d. $ N(0,\sigma^2)$.  By the above definition, the i.i.d. model we studied before can be considered as the special case with order $p$ equal to zero. Next we illustrate the model with $p=1$ : $y_t=\beta y_{t-1}+\epsilon_t$. $AR(p)$ with $p>1$ can be studied with more complicated vector algebra using a multidimensional normal prior. We also assume that the time series has mean zero for simplicity. 

  For technical reasons, in a $AR(1)$ model, we usually assume that $|\beta|<1$, which ensures  that model (\ref{eqn:ar}) admits a unique stationary and causal solution. This constraint is not going to be put on our model, since we will later assume a Gaussian prior for $\beta$, which has infinite support.

The model is same as before, except now we assume that for a segment $[i,j]$, given the hidden parameters $\beta $ and $ \sigma^2$, the observations are generated by
\begin{eqnarray*}
y_i&=&\epsilon_i \\
y_{t}&=&\beta y_{t-1}+\epsilon_{t}, t=i+1,\ldots,j\\
\epsilon_t &\stackrel{iid}{\sim} &N(0,\sigma^2), t=i,\ldots,j
\end{eqnarray*}
We put the following prior on the hidden parameters, same as that used in the basic model (\ref{conj}):
\[ \beta_i|\sigma_i^2,A\sim N(\beta_0,\frac{\sigma_i^2}{k_0}) \]
\[ \sigma_i^2|A\sim Inv-\chi^2(\nu_0,{\sigma_0^2})\]
The probability $p(y_{i:j}|1)$ can be computed analytically like (\ref{integrate}).

From the above discussion, it is evident that our model has great flexibility in modeling the dependency using an embedded Bayesian time series model. The only change we need to make is in computing  $p(y_{i:j}|1)$. Thus if the program is written in separate modules, the programming burden when we change to higher order Markov model is minimal, with only one module to be changed. No other parts of the code need to be modified at all. The model taking into account Markov dependency will not be pursued further in this paper.

\subsection{Repeated observations}\label{sec:mult}
In the above, we apply the segmentation algorithm to one sequence at a time, excluding the EM stage where we may use multiple sequences to estimate the hyperparameters. The extension to multiple sequences is motivated by the application to tiling arrays where the same experiment is performed using multiple slides. The extension to the case where we have multiple sequences with the same underlying changepoints is straightforward. First we obtain the hyperparameters as before (using EM if training data is available, or using the default choice (\ref{autoprior}) separately for each sequence). Thus these hyperparameters are fitted independently for each sequence. The variations between the sequences are reflected in the differences of those hyperparameters. This can be regarded as a normalization method for different arrays. If the arrays have been normalized beforehand, we can use only one set of hyperparameters. Generally, for $m$ replicas, we will have $m$ sets of hyperparameters $\{\Theta^{(r)}\}_{r=1}^m$. Assuming independence of observations between the replicas, the probability of observation can be factored as $p(\{y_{i:j}^{(r)}\}_{r=1}^m|1)=\prod_{r=1}^m p(y_{i:j}^{(r)}|1,\Theta^{(r)})$. The recursions and sampling is same as before, replacing $p(y_{i:j}|1)$ with $p(\{y_{i:j}^{(r)}\}_1^m|1)$.

\section{Examples}
We first demonstrate the effectiveness of the algorithm in detecting a single changepoint. The observations are generated as follows:
\begin{eqnarray*}
&&y_i\stackrel{i.i.d.}{\sim}N(0,\sigma^2),i=1,2,\ldots 200,\\
&&y_i\stackrel{i.i.d.}{\sim}N(\mu,\sigma^2),i=201,\ldots,400, \mu>0
\end{eqnarray*}
First, we set $\sigma^2=1$ and vary $\mu$, smaller $\mu$ makes the changepoint harder to detect. For this simulation, we set the hyperparameters as in (\ref{autoprior}). We generate 100 sequences and draw 100 segmentation samples from the posterior for each sequence and count the number of times the algorithm correctly detect the changepoint. The mean number of times that the posterior sample contains 2 segments averaged over 100 sequences is reported in Table \ref{table:onechangepoint}. For every posterior sample that has 2 segments, we also calculated the distance of the detected changepoint from the true changepoint. From the table, we see that the changepoint model performs reasonably well with $\mu=1$ which is same as the standard deviation of the noise.
\begin{table}
\begin{tabular}{ccccc}
\hline
& $\mu=0.2$ & $\mu=0.5$ &$\mu=1$ &$\mu=2$\\
\hline
mean number of times sampled &&&&\\
segmentation contains 2 segments&44.6& 54.6 &73.2& 99.3\\
\hline
median distance of detected &&&&\\
changepoint from truth &58& 8.5 & 0.6&0.1\\
\hline
\end{tabular}
\medskip
\caption{Number of times sampled segmentation has 2 segments from 100 posterior samples, averaged over 100 sequences, together with the error of the changepoint location averaged over all samples with 2 segments. }
\label{table:onechangepoint}
\end{table}

\begin{figure}[t]
\centering
\begin{tabular}{cc}
\includegraphics[width=7.5cm]{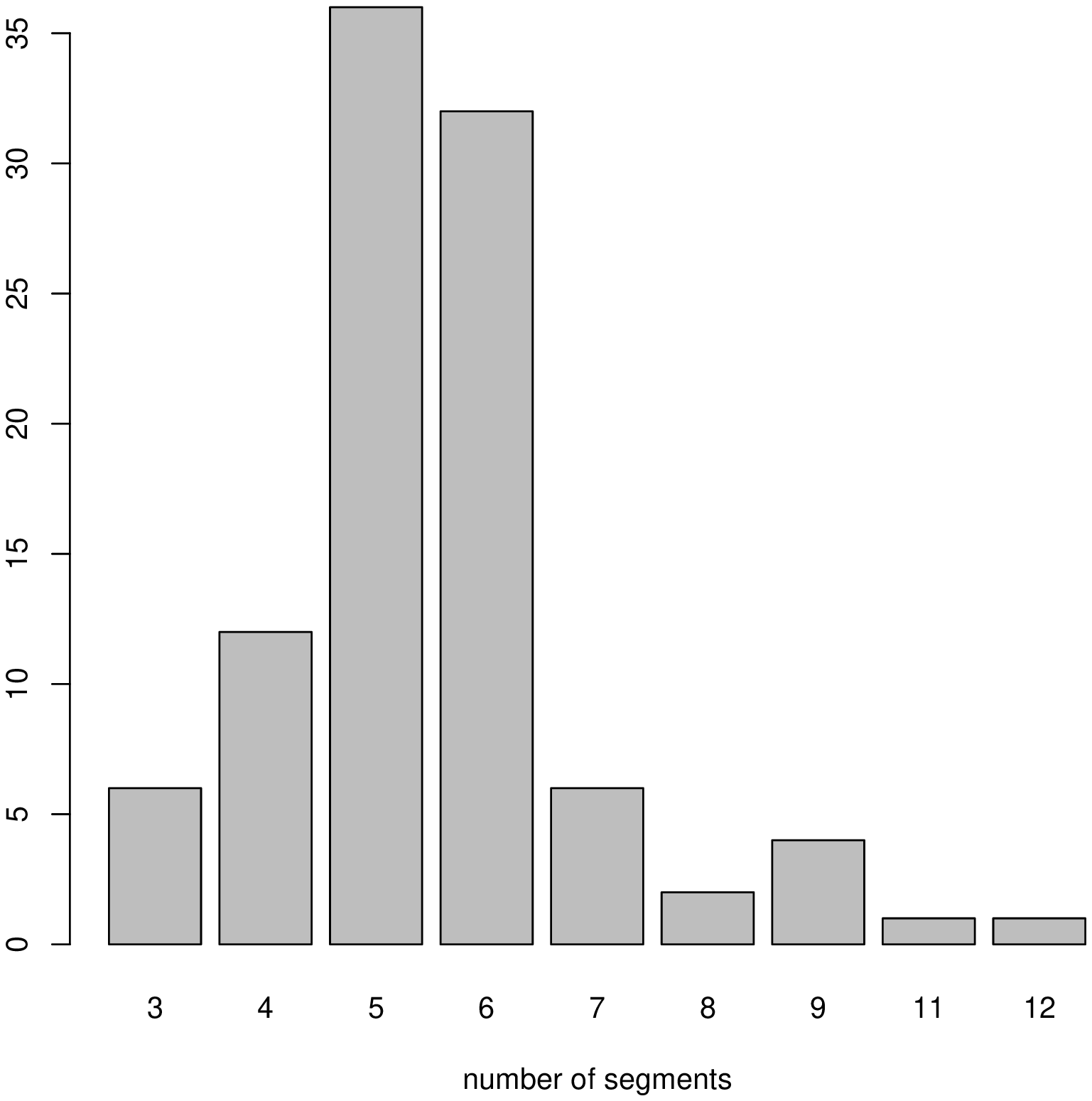}
&\includegraphics[width=7.5cm]{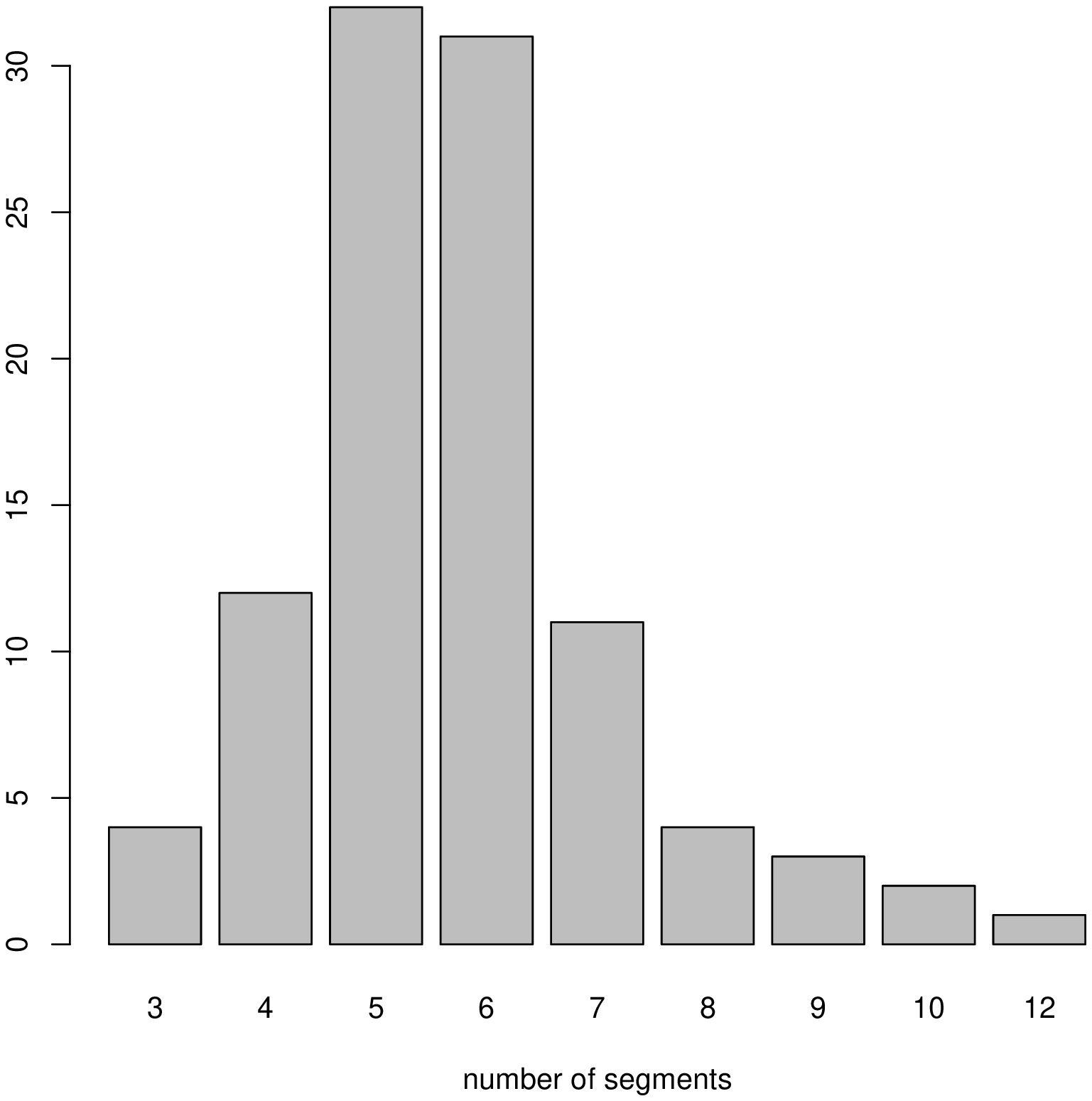} \\
(a)&(b)
\end{tabular}
\caption{Number of segments estimated as the posterior mode using hyperparameters estimated using MCEM (a) or default choice (b), for 100 simulated sequences. }
\label{fig:barplot}
\end{figure}

Next we study multiple changepoint problems and the MCEM algorithm for estimating the hyperparameters. We generate 200 sequences with exactly 5 changepoints, where the observations are generated from our Bayesian model stated in Section \ref{sec:model}. That is, for each sequence, the changepoints are generated from the uniform distribution as in (\ref{aprior}), and the mean and variance for each segment is generated from the conjugate prior with hyperparameters $\mu_0=0, k_0=0.5, \nu_0=5, \sigma_0^2=0.1$. We compare two ways for hyperparameter estimation. For estimation using EM, we use the first 100 sequences for training and the rest 100 sequence for testing the performance of the segmentation algorithm. For estimation using default choice (\ref{autoprior}) of the hyperparameter, we only use the second set of 100 sequences. The MAP estimate of the number of segments for those 100 sequences are shown in Figure \ref{fig:barplot}. Both plots give reasonably good estimate, although the results with estimated hyperparameters seems to be slightly better.

\begin{figure}[t]\label{fig:cgh}
\centering
\includegraphics[width=8cm]{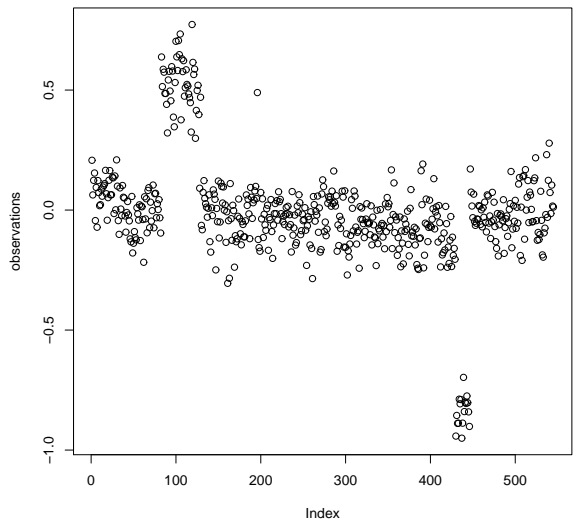}
\caption{CGH array data. }
\end{figure}

Finally, we apply our model to the public CGH array data from \cite{snijders}. CGH arrays are used for the purpose of detecting changes in the number of DNA copies in the samples associated with cancer activity. DNA from a tumor sample and normal reference sample are labeled with different  fluorophores and hybridized to the array. The ratio of the intensity of the tumor to that of the reference DNA is used to measure the copy number changes for a particular location in the genome. We choose to use chromosome 1 through 5 from the cell line gm13330. The missing observations are ignored and we take the observations as equally spaced along the genome. The data from chromosomes 1 through 5 are concatenated and treated as one single sequence, which is plotted in Figure \ref{fig:cgh}. For this problem, we specify the hyperparameters with the default choice (\ref{autoprior}). The segmentation algorithm correctly identifies the visually obvious amplification at position 100 and deletion at around 430. But it does not pick out the single amplified observation at around 200.

\section{Discussion}
We use the simplest prior for the changepoints in this paper. More complicated priors can be specified for the segmentation. For example, we can use a Gamma distribution for the length of each segment (so the distribution on the number of changepoints is implicitly specified), which is a popular choice in the literature of generalized hidden Markov models (i.e. HMM with explicit durations), in application domains such as speech processing (\cite{levinson}). In most applications the main target of inference is the segmentation (i.e. the location of the changepoints) as well as the segmental parameters such as the segmental means, both can easily be obtained from our model. If one considers changepoint model as a tool for function approximation, piece-wise linear model seems to be more appropriate compared to the piece-wise constant model considered here.

In future work, we hope to apply the model on biological experiment with multiple arrays as explained in section
\ref{sec:mult} when replicated data become available. We are currently also investigating combining the hierarchical Bayesian model with HMM so that each segment is assigned a state, which might be useful for some biological applications such as searching for protein binding sites.

\section*{Acknowledgement}
This work represents part of my doctoral dissertation, written at Brown University under the supervision of Professor Charles Lawrence. I owe much to his generous insights and suggestions.





\begin{thebibliography}{99}


\bibitem[Brockwell and Davis(1987)]{brockwell} Brockwell, P.J., Davis, R.A., 1987. Time series: theory and methods, 2nd ed. Springer.

\bibitem[Ding et al.(2005)]{ding} Ding, Y., Chan, C.Y., Lawrence, C.E., 2005. RNA secondary structure prediction by centroids in a Boltzmann weighted ensemble. RNA. 11(8), 1157-1166.

\bibitem[Fan and Yao(2003)]{fan} Fan, J., Yao, Q., 2003. Nonlinear time series: nonparametric and parametric methods, Springer.

\bibitem[Fraley and Raftery (2007)]{rafterymixture}Fraley, C., Raftery, A.E., 2007. Bayesian regularization for normal mixture estimation and model-based clustering. Journal of Classification, to appear.

\bibitem[Gelman et al.(1995)]{gelman} Gelman, A., Carlin, J.B., Stern, H.S., Rubin, D.B., 1995. Bayesian data analysis. Chapman and Hall.

\bibitem[Green(1995)]{green} Green, P., 1995. Reversible jump Markov Chain Monte Carlo computation and Bayesian model determination. Biometrika. 82(4), 711-732.

\bibitem[Levinson(1986)]{levinson} Levinson, S.E., 1986. Continuously variable duration hidden Markov models for automatic speech recognition. Computer Speech and Language. 1, 29-45.

\bibitem[Little and Rubin(2002)]{little} Little, R., Rubin, D., 2002. Statistical analysis with missing data. Wiley.

\bibitem[Liu and Lawrence(1999)]{liu} Liu, J.S., Lawrence, C.E., 1999. Bayesian inference on biopolymer models. Bioinformatics. 15(1), 38-52.

\bibitem[Rabiner(1989)]{rabiner} Rabiner, L.R., 1989. A tutorial on hidden Markov models and selected applications in speech recognition. Proceedings of the IEEE. 77(2), 257-286.

\bibitem[Raftery and Akman(1986)]{raftery} Raftery, A.E., Akman, V.E., 1986. Bayesian analysis of a Poisson process with a changepoint. Biometrika. 73(1), 85-89.

\bibitem[Snijders et al.(2001)]{snijders} Snijders, M.J. et al., 2001. Assembly of microarrays for genome-wide measurement of DNA copy number. Nature Genetics. 29, 263-264.

\bibitem[Wu(1983)]{wu} Wu, C.F.J., 1983. On the convergence properties of the EM algorithm. Annals of Statistics. 11(1), 95-103.

\bibitem[Yang and Kuo(2001)]{yang} Yang, T.Y., Kuo, L., 2001. Bayesian binary segmentation procedure for a Poisson process with multiple changepoints. Journal of Computational and Graphical Statistics. 10(4), 772-785.

\end{thebibliography}
\end{document}